\documentclass[12pt,amsart]{article}
\usepackage{amsmath,amssymb}
\usepackage{bm}
\usepackage[dvipdfmx, hiresbb]{graphicx}
\usepackage{ascmac}
\usepackage{setspace}
\usepackage{theorem}
\theoremstyle{plain}
\theorembodyfont{\upshape} 

\usepackage{lscape}
\usepackage{dcolumn}
\usepackage{longtable}
\usepackage{booktabs,caption}
\usepackage[flushleft]{threeparttable}
\usepackage{multicol}
\usepackage{mdframed}
\usepackage{colortbl}
\usepackage{natbib}
\makeatletter
\def\BibTeX{{\rm B\kern-.05em{\sc i\kern-.025em b}\kern-.08em
    T\kern-.1667em\lower.7ex\hbox{E}\kern-.125emX}}
\makeatother

\usepackage{geometry}
\begin{document}


\onehalfspacing

\title{How Can We Induce More Women to Competitions? \ \\
On the Role of Prosocial Incentives\footnote{We are grateful to our advisors Hidehiko Ichimura and Yasuyuki Sawada for continuous guidance and financial support. We would like to thank Daiji Kawaguchi, Ayako Kondo, Hideo Owan, Katsunori Yamada, and participants of Tokyo Labor Economics Workshop and the seminars at University of Tokyo, Yokohama National University and Rissho University for helpful comments and discussions. All remaining errors are attributable to the authors. }}
\author{\sc{Masayuki Yagasaki\footnote{University of Tokyo, yagamasajp@gmail.com} }\ \ \
\sc{Mitsunosuke Morishita\footnote{University of Tokyo, mitsunosuke0619@gmail.com}} }
\date{This version: \today}

\maketitle

\abstract
Why women avoid participating in a competition and how can we encourage them to participate in it? In this paper, we investigate how social image concerns affect women's decision to compete. We first construct a theoretical model and show that participating in a competition, even under affirmative action policies favoring women, is costly for women under public observability since it deviates from traditional female gender norms, resulting in women's low appearance in competitive environments. We propose and theoretically show that introducing prosocial incentives in the competitive environment is effective and robust to public observability since (i) it induces women who are intrinsically motivated by prosocial incentives to the competitive environment and (ii) it makes participating in a competition not costly for women from social image point of view. We conduct a laboratory experiment where we randomly manipulate the public observability of decisions to compete and test our theoretical predictions. The results of the experiment are  fairly consistent with our theoretical predictions. We suggest that when designing policies to promote gender equality in competitive environments, using prosocial incentives through company philanthropy or other social responsibility policies, either as substitutes or as complements to traditional affirmative action policies, could be promising.

\ \\
\providecommand{\keywords}[1]{\textbf{\textit{Keywords---}} #1}
\providecommand{\jel}[1]{\textbf{\textit{JEL Classification---}} #1}

\keywords{Behavioral economics, Gender, Pro-social incentive, Social image, Competitiveness}

\jel{C91, D82, D91, J16}


\clearpage

\section{Introduction}

While women in many developed countries are attaining higher education at similar rates as men, the gender gap in earnings still exists and women are significantly less likely to hold executive positions in firms. In addition to the more traditional explanations for gender differences in labor market outcomes such as discrimination, the gender differences in abilities and preferences over jobs, another important factor that nowadays economists believe that is important is the gender differences in competitiveness. Seminal work by \citet{NV2007} find that men choose to enter a tournament in the laboratory twice as many as women, despite no difference in performance. A number of experimental researches have replicated the result of \citet{NV2007} and, recently, there is mounting evidence of gender differences in competitiveness measured in the laboratory correlates with the decisions and outcomes in the field \citep[e.g.][]{BNO, RSL2015, Berge2015, Almas2016, Yagasaki}.\footnote{See \citet{Niederle2017} for a survey.} 

Obviously, from an efficiency point of view, it is costly if women with the highest ability do not apply jobs that they are best suited because of its competitive screening process. It is therefore important to ask why women avoid participating in a competition and how can we encourage them to participate in it. 

 Most of the prior research take the gender differences in competitiveness as given and investigated the effect of affirmative action policies favoring women \citep{NSV2013, BS2012}. This strand of research generally show that affirmative action policies such as quotas or preferential treatments are successful in encouraging high performing women to enter competitions and, despite the usual criticism for causing reverse discrimination, there is limited or no efficiency cost.

However, an important point which has been neglected from the prior research is the effect of social image concerns on women's decision to compete. In reality, many of our daily activities are observed by the people around us such as colleagues or classmates. There, it is important for us to think about, at least to some extent, how our behavior will be perceived by others. In fact recent literature of economics has shown surprising degree of influence of such social image concerns in many areas including charitable donation \citep[e.g.][]{Ariely2009, DLM2012}, education \citep[e.g.][]{BJ2015, BEJ2017}, and voting \citep[e.g.][]{Gerber2008}. \footnote{See \citet{BJ2016} for a survey.} Especially, independently from our work, \citet{BFP2017} provide evidence suggesting that single women might hesitate to show characteristics regarded as ``ambitious" or ``acting male" under a public situation because it  hurts them in the dating market. These results suggest that, even if we attempt to encourage women's participation in competitions by introducing affirmative action policies, its effect may be limited to the extent that social image of women participating in competitions departs from traditional female gender norms. 

In this paper, we investigate how social image concerns affect women's decision to compete under three different situations: baseline treatment (no intervention), preferential treatment which is one form of affirmative action policies that preferentially treats competing women, and prosocial incentive treatment under which the donation opportunity is associated to the competitive environment. We first construct a simple theoretical model that incorporates social image concerns with the standard economic incentives when deciding whether to compete and generate predictions. Our model assumes that there are two types of women; female stereotypical types and male stereotypical types. Women with female stereotypical types are assumed to be intrinsically female stereotypical, i.e. have high cost in participating in competition and have high motives to contribute to the society. On the other hand women with male stereotypical types are assumed to be more $manly$, i.e. have lower cost participating in competition and more selfish than women with female stereotypical types. We show that if women are strongly motivated to signal that they are of female stereotypical types, due to the dating market concerns \citep[e.g.][]{Fisman2006, BFP2017} or more psychological such as a direct shame effect of ``acting male", participating in a competition under baseline or affirmative action policies is costly from a social image point of view resulting in low participation rates under public environment. On the other hand, participating in a competition with prosocial incentives is not costly from social image point of view since it does not contradicts with the female stereotypical behavior which suggests women to be socially oriented. Therefore, our model suggests that introducing prosocial incentives to the competitive environment plays two roles: (i) it induces women  who are intrinsically motivated by prosocial incentives to the competitive environment and (ii) it makes participating in a competition not costly for women from social image point of view. 

Based on these theoretical predictions, we conduct a laboratory experiment based on \citet{NV2007} and test these hypothesis.  In our experiment, we randomly assigned sessions into two conditions: ``private" and ``public".  In the private condition, subjects' decision to compete in the experiment were kept completely private as in the standard design. On the other hand in the public condition, subjects were told that their decisions in the experiment would be made public to the subjects participated in the same session at the end of the experiment. We investigate the impacts of this public observability on women's decision to compete under above three different situations: baseline, preferential and prosocial incentive treatments. 

Our experimental results are fairly consistent with the theoretical predictions. We find that both preferential and prosocial incentive treatments are effective in inducing women to competitions in the private condition. However, when decisions to compete are made public, we observe statistically significant declines in women's tournament participation rate in baseline and preferential treatments but not in the prosocial incentive treatment which support our theoretical predictions. On the other hand, we see no statistically significant effect under public observability for men in each treatment. The results are further supported by the regression analysis and robust to additional controls. Overall, we have experimentally identified significant impacts of social image concerns on women's decision to compete and find that prosocial incentive treatment are robust to those concerns.

\subsection{Related Literature}
Our paper adds to at least three strands of research. First, it adds to the literature of gender differences in competitiveness and the effects of policy interventions on women's decision to compete \citep[e.g.][]{NV2007, NSV2013, BS2012}. The present paper contributes to this literature by showing that a social image concern as a significant factor that explains women's underrepresentation in competitive environments. There are two related research that study the effect of other-regarding preferences on  women's decision to compete. \citet{CWZ2016} collected data from parents of middle and high school students in China and show that using child-benefiting voucher as rewards remove the gender differences in competitiveness of those parents. \citet{Samek2015} uses a large field experiment and show that introducing job within prosocial framing increases its application rates. Although these studies are similar in motivations to our paper, they do not consider robustness of the results to public situations which is unique in our study. 

Secondly, the study also has a clear link to the literature of the effects of prosocial incentives on labor market outcomes. There are numerous experimental researches that demonstrated the effectiveness of prosocial incentives on workers effort or performance \citep[e.g.][]{TV2010, TV2015, Imas2014, CCS2016, DP2016}. While most of the previous research show that prosocial incentives are powerful to induce effort or enhance productivity in a given working environment, there are relatively few research studying whether prosocial incentives are enough to motivate workers to self select into such an environment. Indeed, it has been noted by \citet{LMW2012} the importance of self selection in economic experiments, especially in the context of social preferences. Our paper, as well as \citet{CWZ2016} and \citet{Samek2015}, compliments the previous research in that it asks how effective are prosocial incentives in motivating workers to self-select an competitive environment. We show that prosocial incentives are effective and robust to the situations where decisions are publicly observed.

Finally, our paper adds to the recent growing literature of social pressure or social image concerns including charitable donation \citep{Ariely2009, DLM2012}, education \citep[e.g.][]{BJ2015, BEJ2017}, and voting \citep[e.g.][]{Gerber2008}.  The most closely related paper to our paper is \citet{BFP2017}. \citet{BFP2017} find that in a survey to be used by students' career advisers, single, female MBA students under-report their financial ambitions, their willingness to have longer work hours and to travel for work, and some of their personality traits (such as ambition and leadership in day-to-day interactions) when they believe that their classmates will observe their choice. These effects are much weaker among married female students and are not present among male students. Similar to them we also show that, by making use of a controlled laboratory experiment, women avoid participating in competitions, which is generally regarded as a part of male gender norms, when their decisions are observed by others. However, in contrast to \citet{BFP2017}, we move one step further and ask how can we mitigate the negative effect of social pressure or social image concerns on women's decision to compete and show that using prosocial incentives would be a promising approach since it generates positive social image on participating in competitions. To the best of our knowledge, this is the first paper proposing such an idea to mitigate the effects of social pressure.\footnote{See \citet{BJ2016} for a discussion of how one can shape the effects of social pressure. Our idea is not discussed there. }  

The rest of this paper is set out as follows. In Section 2 we construct the theoretical model that incorporates social image concerns when deciding whether to compete and generates theoretical predictions of each treatment. In Section 3 we describe the design of the experiment and state our main testing hypothesis. The results are discussed in Section 4 and Section 5 discusses the policy implication of the results and concludes.

\section{Theoretical Model}
In this section, we construct a simple theoretical model  in order to clarify the role social image concerns plays in women's decision to compete. Various simplifying assumptions imposed on the model are not essential in generating our hypotheses. We first clarify why affirmative action policies such as quotas or preferential treatments, as well as baseline, could be fragile to public observability.

The model below basically follows the framework first proposed in  \citet{BT2006} and is a direct extension of the model in \citet{BEJ2017}. There is also a clear link between our model and the identity model of \citet{AK2000}.  

\subsection{Baseline and Affirmative Action Policies}
\subsubsection{Baseline}
Suppose there is a continuum of women and assume that they have to decide whether to participate in a competition that delivers benefit $b_T$ if they win the competition and zero if not. The winning probability of the competition is denoted as $w \in (0, 1)$. On the other hand, if they decide not to compete, then they receive $b_P$ for sure. We assume that there are two types of women $\tau_i \in \{f, m\}$ which is private information. Here, women of type $f$, referred to as women with female stereotypical types, are interpreted as intrinsically stereotypical women while those whose type is $m$, referred to as women with male stereotypical types, are interpreted as more manly than women with female stereotypical types. We denote the share of women with male stereotypical types by $q$; $Pr(\tau_i=m)=q$. Each of gender types is of two dimensional and $g=(c_g, \theta_g)$ where $g=f, m$. 
The first component $c_g$ is a cost associated with participating in competitions and we will assume that $0=c_m<wb_T-b_P<c_f$. In words, women with male stereotypical types have no cost participating in a competition but women with female stereotypical types have high cost participating in a competition. 
The second component $\theta_g$ represents the degree of women's altruism and we assume that $\theta_m< 0< \theta_f-c_f$. This represents that women with female stereotypical types are more altruistic or socially minded compare to the women with male stereotypical types. \footnote{In this modeling strategy, we assume that preference for competitions and degree of altruism perfectly correlate. 
This is just for a simplification but there are some empirical evidences that support this assumption \citep[e.g.][]{Bartling2009, BKS2012, SS2013}. In the following analysis, we also see that WTP for donation, which is a measure of altruism in our analysis, negatively correlates with tournament entry in the private condition (see Table \ref{cluster}).} If a woman enters the competition, the total payoff is, 
\[w b_T-c_i +\lambda \Pr(\tau_i=f \mid d_i=1),\]
whereas if a woman does not participate in the competition, her total payoff is, 
\[b_P +\lambda \Pr(\tau_i= f \mid d_i=0).\]
The last term represents the benefit of being perceived as female stereotypical types given their choices and $\lambda\geq 0$ captures the strength of such social image concerns. The underlying mechanism that generates this utility could be dating or marriage markets concerns \citep[e.g.][]{Fisman2006, BFP2017} or more psychological such as a direct shame effect of ``acting male".

This setting is completely analogous to \citet{BEJ2017} with minor notational differences. The following proposition corresponds to Proposition 1 in \citet{BEJ2017} and state that, if the decision to compete is publicly observed, women are discouraged to enter into the competitive environment because of the social image concerns. Throughout, we denote $r = \Pr(d_i=1\mid \tau_i=f)$ and $\rho=\Pr(d_i=1\mid \tau_i=m)$. 

\ \\
{\bf Proposition 1.} {\em (i) If $\lambda \leq wb_T-b_P$, $r=0$ and $\rho =1$, (ii) if $\lambda \in \big(wb_T-b_P, \frac{wb_T-b_P}{1-q}\big)$, $r=0$ and $\rho =\frac{1}{q}(1-\frac{\lambda}{wb_T-b_P}(1-q)) \in (0, 1)$ and (iii) if $\lambda \geq \frac{wb_T-b_P}{1-q}$, $r=0$ and $\rho=0$. }

\subsubsection{Affirmative Action Policies Favoring Women}
Consider the same model but with affirmative action policies favoring women. Note that affirmative action policies such as quotas or preferential treatments increase women's winning probability of the competition. Then let $w_A$ be the winning probability under an affirmative action policy and suppose $0=c_m<c_f<w_A b_T-b_P$. This assumption guarantees that the affirmative action policy is effective enough to encourage women with female stereotypical types to participate in a competition under a private situation. However, still in this case, the participation rate converges to zero as $\lambda$ gets larger. 

\ \\
{\bf Propostion 2.} {\em (i) if  $0 \leq \lambda \leq \frac{w_Ab_T-b_P-c_f}{q}$, $r=1$ and $\rho = 1$, (ii) if $w_Ab_T-b_P-c_f \leq \lambda \leq \frac{w_Ab_T-b_P-c_f}{q}$, $r=\frac{q}{1-q}\frac{\lambda-(w_Ab_T-b_P-c_f)}{w_Ab_T-b_P-c_f}$ and $\rho=1$, (iii) if $w_A b_T-b_P-c_f \leq \lambda \leq w_Ab_T-b_P$, $r=0$ and $\rho =1$, (iv) if $w_Ab_T-b_P \leq \lambda \leq \frac{w_Ab_T-b_P}{1-q}$, $r=0$ and $\rho = \frac{1}{q}(1-\frac{\lambda}{w_A b_T-b_P}(1-q))$, (v) if $\frac{w_Ab_T-b_P}{1-q}\leq \lambda$, $r=0$ and $\rho = 0$. } \footnote{Note that the equilibrium in (ii) is unstable with respect to perturbation in $\lambda$. }

      \subsection{Introducing Prosocial Incentives}
 The reason why baseline and affirmative action policies were fragile to public observability was that women with female stereotypical types always had a weaker incentive  to participate in the competition compare to women with male stereotypical types. This made the noncompetitive environment always beneficial from social image point of view and  uniquely supported the equilibrium with no participation for $\lambda$ sufficiently large by D1 criterion \citep{CK1987, BS1987}. 
 
The idea here is to use prosocial incentives to reverse this relationship. We expect introducing prosocial incentives in the competitive environment plays two roles: (i) it induces women with female stereotypical types, who are intrinsically motivated by prosocial incentives, to a competition and (ii) it makes participating in a competition not costly for women since  women with female stereotypical types are now having stronger incentive to participating in it. In contrast to affirmative action policies, (ii) makes the effect of introducing prosocial incentives on women's decision to compete robust to the public observability. 
 
 Suppose if women participate in a competition, in addition to financial reward $b_T$, they are given an opportunity to make charitable donation if they win. Denote $a_i \in \{0, 1\}$ a discrete decision whether to make charitable donation or not. Then, the payoff obtained by participating in the competition is given by, 
 \begin{eqnarray}\nonumber
           \max_{a_i \in \{0, 1\}} w(b_T+\theta_i a_i)-c_i + \lambda \Pr(\tau_i=f\mid a_i, d_i=1),
 \end{eqnarray}      
  whereas the payoff of not participating in the competition is the same with the previous cases. Since we are assuming $\theta_m < 0<\theta_f-c_f$, note that now women with female stereotypical types always have a stronger incentive to participate in a competition compare to the women with male stereotypical types. 
      
  The following Proposition 3 shows that, indeed, this treatment accomplishes full participation for any $\lambda \geq 0$. The only decision that $\lambda$ matters is the decision to donate. Throughout we denote $r_T=\Pr(a_i=1\mid  \tau_i=f, d_i=1)$ and $\rho_T=\Pr(a_i=1\mid \tau_i=m, d_i=1)$. 
 
\ \\
{\bf Proposition 3.} {\em For every $ \lambda \geq 0$, $r=1$ and $\rho=1$. In addition, (i) if   $0\leq \lambda \leq -w\theta_m$, $r_T=1$ and $\rho_T=0$, (ii) if $-w\theta_m \leq \lambda \leq -\frac{w\theta_m}{1-q}$, $r_T=1$ and $\rho_T=1-\frac{1}{q}(1+\lambda \frac{1-q}{w\theta_m})$, (iii) if $\lambda \geq -\frac{w\theta_m}{1-q}$, then $r_T=1$ and $\rho_T=1$.
}

\ \\ 
Finally, before moving on, it is worthwhile to mention the case of men. For men, all of the previous analysis can be easily adjusted by assuming that men benefit from being perceived by others as male stereotypical types given their choices. In this case, the model predicts that men are encouraged to enter into the competitive environment when the decision is publicly observed by others so long as $\lambda$ is large. However, in the previous literature, there is evidence that men are not so sensitive to social image concerns compare to women and public observability has little impact on behaviors for men \citet[e.g.][]{Ariely2009, BFP2017}. Therefore, we expect that the effect of public observability is small or approximately zero (i.e., $\lambda=0$) in contrast to women and mainly focus on women in the following analysis.

\section{Experimental Procedures}
The experiment was conducted in November, December of 2016. We conducted 12 sessions in total, 10 at the University of Tokyo, 2 at Keio University. Our subjects were drawn from students in the University of Tokyo, Keio University, Waseda University. Subjects were recruited through fliers posted throughout the campus, handed out and uploaded on Twitter and Facebook by our research assistants. We randomly chose 150 subjects for both men and women from the students who were interested in our research and had registered at an online web site. Finally 279 subjects (138 men, 141 women) participated in the experiment. In this paper, however, we only utilize the sample of students in the University of Tokyo. All those students were participating in the sessions conducted at the University of Tokyo. This leaves us with a sample of 188 subjects (97 men, 91 women). \footnote{ In total, subjects were 188 from the University of Tokyo, 69 from Keio University, 22 from Waseda University. We omit 91 data from Keio and Waseda University. The results do not change qualitatively even if we include those samples.}
  
 Upon arrival, subjects were told that they would receive a participation fee, 4000 yen, and that they might earn additional money, 2000 yen on average, depending on their performance in the experiment. \footnote{They were also informed that they had to come to the University of Tokyo or Keio University later to receive their total fee. Our participation fee is relatively high compared to the literature since it includes the transportation cost during the experiment. }
  
  Our experimental design basically follows the standard experimental design as in \citet{NV2007}. At the start of the experiment, subjects were told they were randomly divided into groups of 3 men and 3 women 
 and that they would be performing a number of tasks. One of these would be randomly chosen for payment at the end of the experiment. In each round subjects had 3 min to solve as many as possible mazes as in Figure \ref{fig:maze} below.

\begin{figure}[h]
 \begin{center}
  \includegraphics[width=75mm]{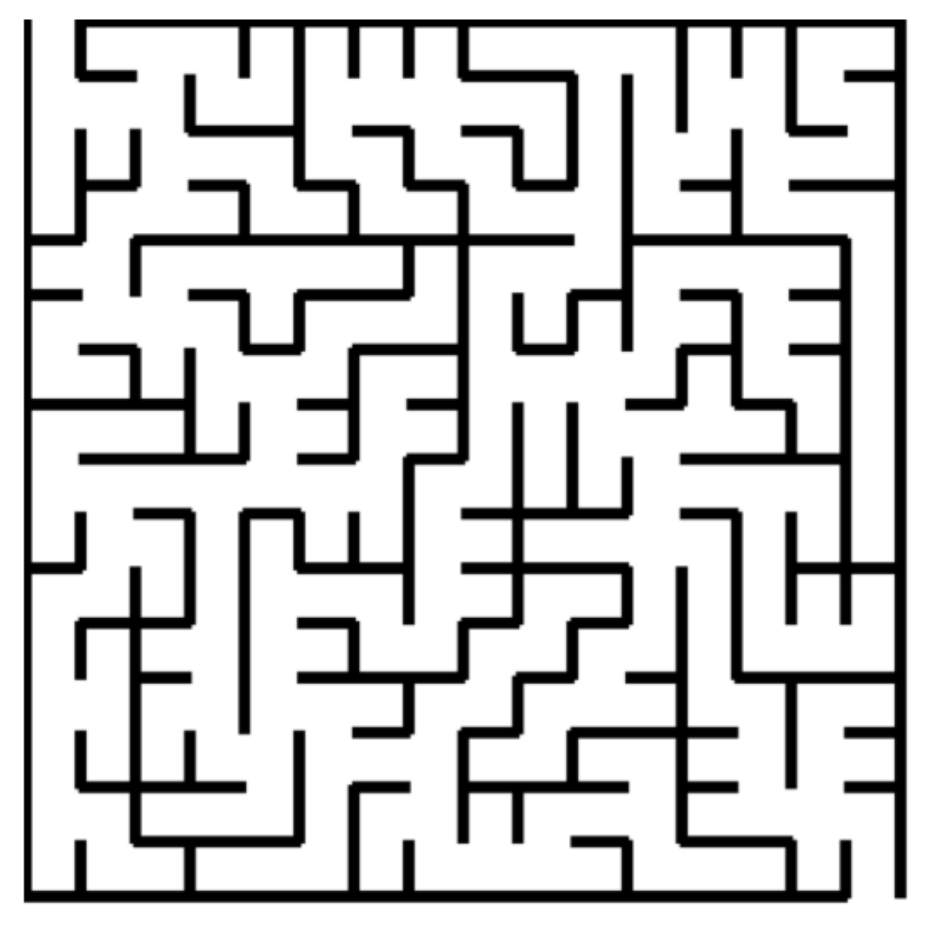}
 \end{center}
 \caption{an example of maze}
 \label{fig:maze}
\end{figure}

 Before starting the first task, subjects were given an instruction of how to solve mazes, and given an opportunity of 3 min practice. Subjects were informed of the nature of the task to be carried out and the payment scheme only immediately before performing the task. The specific payment schemes of the first two rounds are the followings. 
 
 \ \\
 {\bf Round 1: Piece Rate.} Subjects were asked to solve as many mazes as possible in 3 min. They would receive 50 yen for each maze correctly solved if this task was randomly selected for payment.
 
 \ \\
 {\bf Round 2: Compulsory Tournament.} Subjects were asked to solve as many mazes as possible in 3 min. If this task was randomly selected for payment, the two subjects who solved the largest number of correct mazes in
the group each would receive 150 yen for each maze correctly solved while the other subjects would receive no payment. In case of ties the winner were chosen randomly.
 
 \ \\
 In rounds 3 to 5, in order to asses the impacts of different compensation schemes, we use a within-subjects design. Subjects sequentially faced with the following treatments, with the treatment order was randomly assigned to rounds 3 to 5 to minimize order effects, and had to choose under which payment scheme their next performance would be compensated. 
 
 \ \\
 {\bf Baseline Treatment. } Subjects were asked to choose either piece rate or tournament and then solve as many mazes as possible in 3 min. Payment would depend on which payment scheme they chose if this round was randomly selected for payment. If a subject were to choose piece rate, she would get 50 yen per maze solved correctly. If she chose tournament, she would get 150 yen per maze solved correctly if her score exceeded that of at least 4 of the other group members in round 2, otherwise she would receive no payment. In case of ties the winner were chosen randomly.  
 
 \ \\
 {\bf Preferential Treatment.} In preferential treatment, each woman's performance was automatically increased by one unit (i.e., one correct maze). Given the augmented scoring rule for women, subjects were asked to choose either piece rate or tournament and then solve as many mazes as possible in 3 min. Payment were determined according to the rules of the baseline treatment. 
 
 \ \\
 {\bf Prosocial Incentive Treatment.} In prosocial incentive treatment, subjects were allowed to donate from their rewards to a charity organization {\em only when they chose tournament.} Subjects were asked to choose either piece rate or tournament and then solve as many mazes as possible in 3 min. In addition, if they were to choose tournament, they were told that they could decide how much of a 150 yen per one correct maze if they won to keep for themselves and how much to pass on to a charity of their choice. 
 
\ \\
In addition to the above standard design that follows \citet{NV2007}, we randomly manipulate the publicly observability of decisions during the experiment to see how social image concerns affect the decisions to compete. We randomly assigned $private$ and $public$ conditions to sessions. 

\ \\
{\bf Private Condition.} Subjects decisions were kept private throughout the experiment as in the standard literature. 

\ \\
{\bf Public Condition.} Subjects were told that there decisions to choose piece rate or tournament would be made public to subjects participating in the same session. In prosocial incentive treatment, in addition, subjects were told that the amount shared to the charity would be also made public if they chose tournament.    

\ \\
In summary, our experimental design is a combination of within-subject design, which is designed to see how each treatment affects the decisions, and between-subject design, which is designed to see how public observability affects the decisions and to explore the robustness of effects of each treatment.  

\ \\
{\bf Questionnaires.} At the end of the experiment, subjects were asked to answer the post-experimental questionnaire. As well as demographic and socio-economic questions, questionnaires include subjects' guess of their relative rank in compulsory tournament, both subjective and objective risk attitude and willingness to pay (WTP) for donation. WTP for donation is used to estimate how much each subject valued the donation opportunity in the prosocial incentive treatment and interpreted as a measure of altruism in our analysis.    
 
\subsection{Descriptive Statistics}
We will be examining tournament entry choice and see whether making the decisions public has an effect or not. In doing so, we wish to control for ability, confidence, risk attitudes and the level of altruism that could affect the decisions of tournament entry. Table \ref{summary_stat_m} and \ref{summary_stat_f} show the summary statistics by gender.

There are no statistical differences between subjects in public and private conditions for most of the variables but we see a few important differences between these groups. For men, subjects in the public treatment are significantly more risk seeking, in the sense that they choose more risky lotteries, than those in the private condition (see Table \ref{summary_stat_m}). On the other hand, for women, subjects in the public condition are less confident, have higher WTP for donation (see Table \ref{summary_stat_f}). These observable differences are important in our analysis since these differences itself could generate differences in tournament entry choice between subjects in public and private conditions.\footnote{The observed imbalance between subjects in public and private conditions may itself be caused by making the decisions public in the public condition. Indeed, in the literature of self-presentation in social psychology, there is an evidence in which behaviors performed under public observability change the self-concept of individuals and result in the change in behaviors (see \citet{Tice1992}). If that is the case, variables which are not balanced should be omitted from the analysis since those are outcomes of the treatment. Therefore, in the following analysis, we assume that this is not the case and use those variables as controls.} Therefore, we will control for these differences in the analysis below.

 \subsection{Testing Hypothesis}
Based on the previous theoretical analysis, we state the following testing hypothesis regarding the effects of each intervention and the public observability on women's decision to compete. First, we expect both preferential treatment and prosocial incentive treatment are effective in inducing women to competitive environment when decisions are completely private. 

\ \\
{\bf Hypothesis 1.} {\em When decisions are private, we expect women's higher tournament participation rate in both preferential and prosocial incentive treatment than in the baseline treatment.} 

\ \\
Our model also predicts that making the decision to compete public would discourage women's tournament participation in baseline and preferential treatments. On the other hand, we expect there is no such an effect in prosocial incentive treatment and thus prosocial incentive treatment is robust to public observability. 

\ \\
{\bf Hypothesis 2.}

{\em (i) When decisions are public, we expect women's lower tournament participation rate in both baseline and preferential treatment than in the respective private case. 

(ii) We expect no significant difference in women's tournament participation rate in prosocial incentive treatment. }

\section{Results}

In this section we will present our experimental results and discuss whether the results are consistent to the stated testing hypothesis. Our preliminary results are displayed in Figure \ref{te} which represents the proportion of subjects who choose tournament in each treatment. 

First of all, in the baseline, although the difference is not statistically significant (tournament entry rate among women and men, respectively: 0.46, 0.57, $p$= 0.30), we observe that women are 11 percentage points less likely to enter into the tournament in the private condition. In the public condition, however, there is a significant gender difference in tournament entry rate (0.24, 0.65, $p=0.000$). Compared to the private condition, women entering into the tournament significantly decreased from $0.46$ to $0.24$ ($p=0.034$), which is consistent to hypothesis 2(i), while men increased their entering into the tournament from 0.57 to 0.65 but not significant ($p=0.416$). 

Secondly, in the private condition, the introduction of preferential treatment increased women entering the tournament from 0.46 to 0.63 and decreased men from 0.57 to 0.41 and the impacts are statistically significant for both women and men ($p=0.032$ and $p=0.034$ for women and men respectively). \footnote{$p$-values are calculated by regressing tournament entry dummy on the dummies of preferential treatment and prosocial incentive treatment clustered by individuals.} This is consistent to hypothesis 1 and as expected since the introduction of preferential treatment increases the probability of winning the tournament of women (and decreases that of men). In the public condition, we see similar trends and tournament entry rate increased for women from 0.24 to 0.51 and decreased for men from 0.65 to 0.53 and the impacts are statistically significant ($p=0.010$ and $p=0.010$). In particular, compared to the private condition, women are still  underrepresented in the tournament and the difference is significant ($p=0.047$) which supports hypothesis 2(i).

Finally, as displayed in Figure \ref{te}, we observe that prosocial incentive treatment is as effective as preferential treatment for women while keeping men's tournament participation more or less similar to that of the baseline in the private condition. Indeed, the actual rate of tournament entry in prosocial incentive treatment are 0.61 for women and 0.50 for men and the impacts, compared with the baseline, are statistically significant for women and not for men ($p=0.092$ and $p=0.326$) which is consistent to hypothesis 1. On the other hand, in the public condition, women dramatically increased their tournament entry rate from 0.24 to 0.57 ($p=0.00$) but not for men. Compared to the private condition, women are no longer underrepresented in the tournament and the difference between private and public conditions is now insignificant ($p=0.767$). This supports our hypothesis 2(ii). 

Overall, the preliminary results are fairly consistent to our testing hypothesis. However, recall that even though we randomized public and private conditions, we had some observable differences between those groups. For example, in our sample, women in the public condition had high willingness to pay for donation compared to women in the private condition. As in the previous theoretically analysis, if women's willingness to compete and willingness to pay for donation are negatively correlated, the results regarding public observability could be totally explained by the negative correlations between those two factors, challenging the validity of hypothesis 2. To avoid these concerns, next we test our hypothesis using regressions controlling for those observable characteristics.

\subsection{Regression Analysis}
 In Table \ref{cluster} we present the results of regressions of women's decision to enter a tournament. \footnote{We had to drop 4 observations since WTP for donation were missing for those observations.}In column (1) and (3), we restrict our sample to women in private and public conditions respectively and regress tournament entry dummy on dummies of preferential and prosocial incentive treatments, probability of winning a tournament\footnote{Following \citet{NSV2013}, for any given performance level at round 2, we draw 1,000,000 groups consisting of 3 men and 3 women using our sample with replacement and calculated the frequency of wins in this set of simulated groups.}, guessed rank and risk attitudes. Standard errors are clustered by individuals to account for the lack of independence among the three individual observations. For preferential treatment, the effect is positive and statistically significant even accounting for the change in winning probability for both private and public conditions. For prosocial incentive treatment, we see that the effect is not statistically significant  but it is still substantial in the private condition. On the other hand, the effect of prosocial incentive treatment is strong under public observability. 
 
 In column (2) and (4) we include WTP for donation and an interaction term between WTP for donation and prosocial incentive treatment to explore the mechanism underlying the effect of prosocial incentive treatment. Namely, the introduction of the donation opportunity in the tournament environment changes the benefit associated with entering tournament especially for women with high WTP for donation. If the effect of prosocial incentive treatment is solely driven by women with high WTP for donation, we should observe that the coefficient on interaction term between WTP and prosocial incentive treatment dummy to be positive and the coefficient on prosocial incentive treatment dummy to be zero in column (2) and (4). 
 
 In the private condition, this is actually the case and is indicated in Table \ref{cluster}. Therefore, we conclude that the effect of prosocial incentive treatment in the private condition is driven by inducing women who are intrinsically motivated by the donation opportunity which is consistent to our theoretical analysis. In addition, in the private condition, the coefficient on WTP is negative and significant. This is also consistent to our theoretical model in which we assumed willingness to compete and the level of altruism negatively correlate. 
 
 On the other hand, in column (4), we see that adding WTPs do not reduce the coefficient on prosocial incentive treatment and the effect is still strongly positive and significant under public observability. Indeed, neither WTP nor the interaction term is significant in column (4). This indicates, under public observability, women are motivated to enter the tournament in the prosocial incentive treatment not necessarily because they are intrinsically motivated by the donation opportunity but maybe due to other factors. This is consistent to social image explanation of our theoretical model, i.e., the introduction of prosocial incentives makes entering the tournament not costly from social image point of view. It also supports our statement that the observed strong effect of prosocial incentive treatment under public observability is $not$ driven by having more women with high WTP in the public condition than in the private condition. Overall, the results in Table \ref{cluster} are on line with hypothesis 1 and 2. 
   
To see the effect of public observability on each treatment, Table \ref{public} reports the results of regressions of decision to enter a tournament on the dummy of public observability by each treatment.\footnote{The results do not change even if we use regressions with pooled samples standard errors clustered by individuals.} All specifications include session fixed effects and from column (3) to (6) we include probability of winning, guessed rank and WTP for donation. Column (5) and (6) include demographic controls such as age, mother and father's education and major fixed effects. In the baseline and preferential treatment, the effect of public observability on tournament entry is negatively significant across all specifications for women. The only treatment of which the coefficient on public observability is not significant for women is the prosocial incentive treatment. Altogether, hypothesis 2 is supported. Finally, for men the coefficients are statistically insignificant across all treatments and specifications. This supports our presumption that men are not sensitive to public observability compared to women and is consistent with the previous literature. Table \ref{public} reports $p-$values for gender differences in the coefficients of public observability. We see significant gender differences in the coefficients on public observability of baseline and preferential treatment but not for prosocial incentive treatment which is again consistent to our hypothesis. \footnote{Our theory also predicts that the amount shared to the charity in prosocial incentive treatment will increase under public observability. We also test this prediction. The result partially supported our prediction but it was not  conclusive. We leave this point for future investigation.}

\section{Conclusion}
Taking a society where the social image associated with women participating in a competition contradicts with its cultural female gender norms at face value, it is important to explore policies that may mitigate negative effects of social image concerns or social pressure on women's decision to compete. In this paper, we have demonstrated that social image concerns negatively affect women's decision to compete and that affirmative action policies favoring women are not robust to such an effect. On the other hand, we show that introducing prosocial incentives in the competitive environment is effective since (i) it induces women who are intrinsically motivated by prosocial incentives to the competitive environment and (ii) it makes participating in a competition not costly  for women from social image point of view. Overall, we suggest that when designing policies to promote gender equality in competitive environments, using prosocial incentives through company philanthropy or other social responsibility policies, either as substitutes or as complements to traditional affirmative action policies, could be promising.

An important direction for future research is to extend our findings to investigate various forms of prosocial incentives as in \citet{TV2010, TV2015}, which may contribute to the further understandings of policy design that firms can implement in order to encourage women to enter competitions. Relatedly, aside from prosocial incentives, it might be possible to mitigate negative social image effects through other channels as discussed in \citet{BJ2016}. We plan to study these questions in future projects. 

Finally, future work should address an extremely important open question of how policies that are intended to encourage women to participate in a competition affects the evolution of gender norms or the notion of gender stereotypes in the society in the long-run. One needs to be aware of, for instance, the possibility that affirmative action policies which preferentially treat women can generate a misconception that women are ``weaker", ``less intelligent", and ``inferior" than men, which could mislead to gender stereotypes in an undesirable way (e.g. \citet{Levinson2011}). An ideal policy should be the one which is effective enough in inducing women to competitions and that breaks the glass ceiling faced by women in the long-run. 

\clearpage
\bibliographystyle{aer}
\ifx\undefined\bysame
\newcommand{\bysame}{\leavevmode\hbox to\leftmargin{\hrulefill\,\,}}
\fi


\newpage

\appendix

\section{Tables and Figures}
\begin{table}[!h] 
\caption{Descriptive Statistics of Men by Private and Public Conditions}  
\begin{center}
\begin{tabular}{@{\extracolsep{5pt}}lccccccc}
\\[-1.8ex]\hline 
\hline \\[-1.8ex] 
\\[-1.8ex]
 &\multicolumn{3}{c}{Private} &\multicolumn{3}{c}{Public}& $p$-value\\
  \cline{2-4} \cline{5-7}  \\ 
  & N & Mean & SD & N & Mean & SD & $t$-test \\
 \hline\\[-1.8ex] 
Performance (Piece-rate) & 46 & 10 & 2.4 & 51 & 11 & 3.4 & 0.37 \\
Performance (Tournament) & 46 & 9 & 2 & 51 & 9.4 & 2.5 & 0.39 \\
Performance (B) & 46 & 9.8 & 2.5 & 51 & 9.6 & 2.9 & 0.72 \\
Performance (PRE) & 46 & 10 & 2.8 & 51 & 10 & 2.9 & 0.95 \\
Performance (PRO) & 46 & 11 & 3.1 & 51 & 10 & 3.4 & 0.59 \\
Guessed Rank & 46 & 2.9 & 1.1 & 51 & 2.9 & 1.1 & 0.89 \\
Overconfidence & 46 & 0.72 & 1.9 & 51 & 0.31 & 1.7 & 0.27 \\
Risk (Self-report) & 46 & 4.8 & 2.1 & 51 & 5 & 2.5 & 0.67 \\
Risk (Lottery) & 46 & 3.3 & 1.5 & 51 & 3.9 & 1.5 & 0.046 \\
WTP for Donation & 45 & 2.1 & 1.7 & 50 & 1.8 & 1.2 & 0.37 \\
\\
[-1.8ex]\hline 
\hline \\[-1.8ex] 
\\[-1.8ex]

\end{tabular}

\end{center}
\label{summary_stat_m}
\end{table}

\begin{table}[!h] 
\caption{Descriptive Statistics of Women by Private and Public Conditions}  
\begin{center}
\begin{tabular}{@{\extracolsep{5pt}}lccccccc}
\\[-1.8ex]\hline 
\hline \\[-1.8ex] 
\\[-1.8ex]
 &\multicolumn{3}{c}{Private} &\multicolumn{3}{c}{Public}& $p$-value\\
  \cline{2-4} \cline{5-7}  \\ 
  & N & Mean & SD & N & Mean & SD & $t$-test \\
 \hline\\[-1.8ex] 
Performance (Piece-rate) & 46 & 9.6 & 2.9 & 45 & 9.8 & 2.7 & 0.72 \\
Performance (Tournament) & 46 & 8.9 & 2.3 & 45 & 8.7 & 2.3 & 0.61 \\
Performance (B) & 46 & 9.4 & 2.7 & 45 & 9.6 & 2.6 & 0.7 \\
Performance (PRE) & 46 & 10 & 3.1 & 45 & 9.3 & 2.4 & 0.23 \\
Performance (PRO) & 46 & 10 & 3.6 & 45 & 10 & 3.1 & 0.66 \\
Guessed Rank & 46 & 3.2 & 1.1 & 45 & 3.5 & 1.1 & 0.095 \\
Overconfidence & 46 & 0.28 & 1.6 & 45 & 0.13 & 1.6 & 0.66 \\
Risk (Self-report) & 46 & 4.4 & 2.2 & 45 & 4.2 & 2.2 & 0.78 \\
Risk (Lottery) & 46 & 2.8 & 1.5 & 45 & 2.8 & 1.4 & 0.88 \\
WTP for Donation & 44 & 2.3 & 1.7 & 43 & 3.4 & 1.8 & 0.0043 \\
\\
[-1.8ex]\hline 
\hline \\[-1.8ex] 
\\[-1.8ex]

\end{tabular}

\end{center}
\label{summary_stat_f}
\end{table}

\begin{table}
\caption{The Effects of Preferential and Prosocial Incentive Treatments of Women}
\label{cluster}
\begin{center}
\scalebox{0.9}{
{
\def\sym#1{\ifmmode^{#1}\else\(^{#1}\)\fi}
\begin{tabular}{l*{4}{c}}
\hline\hline
                    &\multicolumn{1}{c}{(1)}&\multicolumn{1}{c}{(2)}&\multicolumn{1}{c}{(3)}&\multicolumn{1}{c}{(4)}\\
                    &\multicolumn{1}{c}{private}&\multicolumn{1}{c}{private}&\multicolumn{1}{c}{public}&\multicolumn{1}{c}{public}\\
\hline
PRE                 &       0.161*  &       0.160*  &       0.196***&       0.194***\\
                    &     (0.085)   &     (0.086)   &     (0.068)   &     (0.069)   \\
[1em]
PRO                 &       0.136   &      -0.037   &       0.326***&       0.305** \\
                    &     (0.094)   &     (0.096)   &     (0.075)   &     (0.120)   \\
[1em]
Performance (Tournament)          &       0.186   &       0.185   &      -0.065   &      -0.053   \\
                    &     (0.153)   &     (0.155)   &     (0.122)   &     (0.123)   \\
[1em]
Guessed rank        &      -0.021   &      -0.021   &      -0.135** &      -0.128** \\
                    &     (0.056)   &     (0.057)   &     (0.052)   &     (0.053)   \\
[1em]
Risk(Self-report)   &       0.055***&       0.053** &       0.070** &       0.069** \\
                    &     (0.020)   &     (0.020)   &     (0.028)   &     (0.028)   \\
[1em]
Risk(Lottery)       &       0.058   &       0.057   &       0.025   &       0.019   \\
                    &     (0.040)   &     (0.041)   &     (0.036)   &     (0.038)   \\
[1em]
WTP for Donation                &               &      -0.056** &               &       0.018   \\
                    &               &     (0.024)   &               &     (0.027)   \\
[1em]
PRO×WTP            &               &       0.132***&               &       0.009   \\
                    &               &     (0.041)   &               &     (0.037)   \\
\hline
Observations        &         132   &         132   &         129   &         129   \\
\hline\hline
\end{tabular}
}

}
\end{center}
       \begin{tablenotes}
      \small
      \item {\footnotesize $Notes.$ The table reports regression results of tournament entry of women by private and public conditions. PRE is the dummy of the preferential treatment. PRO is the dummy of the prosocial incentive treatment. Performance is the probability of winning the round 2 tournament. All specifications include session fixed effects. Standard errors clustered by individuals are in parentheses; * $p<0.1$, ** $p<0.05$, *** $p<0.01$.}
          \end{tablenotes}   
\end{table}

\begin{table}
\caption{The Effects of Public Obserbavility by Gender}
\label{public}
\begin{center}
\scalebox{0.9}{
{ 
 \begin{tabular}{lcccccc}
    \midrule
    \midrule
          & (1)   & (2)   & (3)   & (4)   & (5)   & (6) \\
          & men   & women & men   & women & men   & women \\
         \hline
    \textbf{Baseline} & 0.068 & -0.225** & 0.083 & -0.185** & 0.117 & -0.231*** \\
          & (0.099) & (0.098) & (0.089) & (0.093) & (0.094) & (0.093) \\
    $p$-values for gender differences & \multicolumn{2}{c}{0.032} & \multicolumn{2}{c}{0.029} & \multicolumn{2}{c}{0.004} \\
    \hline
    \textbf{Preferential Treatment} & 0.076 & -0.212** & 0.007 & -0.196* & 0.063 & -0.192* \\
          & (0.103) & (0.104) & (0.097) & (0.106) & (0.102) & (0.110) \\
    $p$-values for gender differences & \multicolumn{2}{c}{0.045} & \multicolumn{2}{c}{0.138} & \multicolumn{2}{c}{0.063} \\
    \hline
    \textbf{Prosocial Incentive Treatment} & 0.076 & -0.042 & 0.077 & -0.073 & 0.099 & -0.073 \\
          & (0.103) & (0.101) & (0.097) & (0.105) & (0.104) & (0.110) \\
    $p$-values for gender differences & \multicolumn{2}{c}{0.403} & \multicolumn{2}{c}{0.273} & \multicolumn{2}{c}{0.215} \\
    \hline
    Session FE & $\surd$     & $\surd$      & $\surd$      & $\surd$      & $\surd$      & $\surd$  \\
    Performance and Psychological Attributes &       &       & $\surd$      & $\surd$      & $\surd$      & $\surd$  \\
    Demographic Controls &       &       &       &       & $\surd$      & $\surd$  \\
    \hline
    Observations & 97    & 91    & 94    & 86    & 93    & 84 \\
    \midrule
    \midrule
    \end{tabular}
}
}
\end{center}
       \begin{tablenotes}
      \small
      \item {\footnotesize $Notes.$ The table reports regression results on the effect of public observability on tournament entry. Performance is the probability of winning the round 2 tournament and psychological attributes include guessed rank, risk attitudes and WTP for donation. Demographic controls include age, mother and father’s education and major fixed effects. All specifications include session fixed effects. Robust standard errors in parentheses; * $p<0.1$, ** $p<0.05$, *** $p<0.01$.}
          \end{tablenotes}   
\end{table}

\clearpage

\begin{figure}[h]
\begin{center}
\includegraphics[width=160mm]{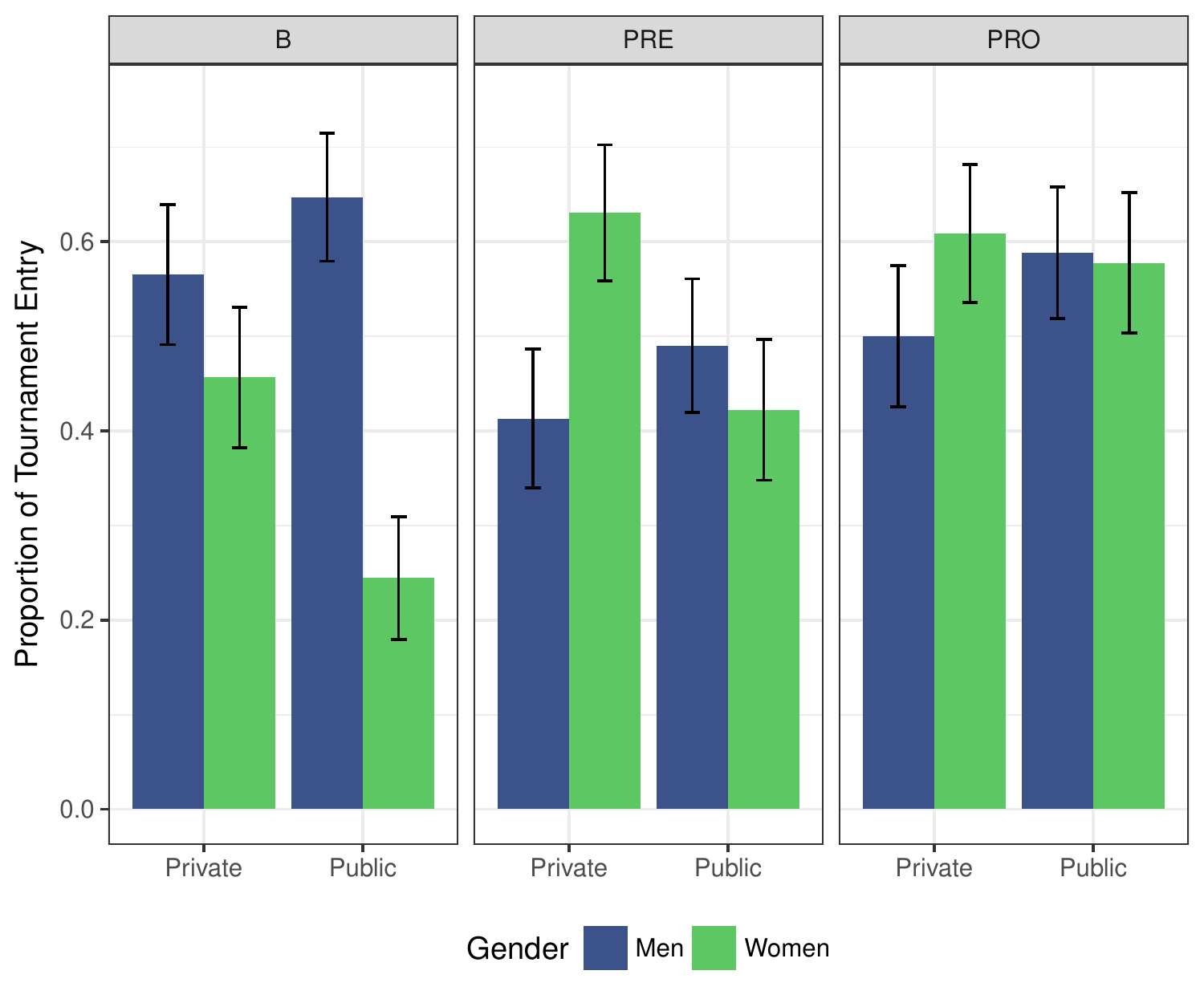}
\caption{Proportion of Subjects Entering Tournaments}
\label{te}
\end{center} 
\end{figure}

\clearpage
\section{Proof of Propositions (Not for Publication)}
$Proof\ of\ Proposition\ 1.$ See \citet{BEJ2017} Proposition 1.   
  
\ \\
$Proof\ of\ Proposition\ 2.$ (i) Suppose $r=0$ and $\rho=0$. Then, $\Pr(\tau_i=f\mid d_i=0)=1$, since $\tau_i=m$ is eliminated by applying the D1 criterion.
Hence, 
\begin{eqnarray}\nonumber
b_P+\lambda  &\leq& w_Ab_T-c_f+\lambda (1-q)  \Leftrightarrow  0 \leq \lambda \leq \frac{w_Ab_T-b_P-c_f}{q}.          
\end{eqnarray}
 (ii) Suppose $r=0$ and $\rho = 1$. Then we have
    \begin{eqnarray}\nonumber
                      &\Pr(\tau_i=f \mid  d_i=0)=\frac{(1-q)(1-r)}{(1-q)(1-r)+q(1-\rho)} =1, \ \\ \nonumber
                        & \Pr(\tau_i=f \mid d_i=1) = \frac{(1-q)r}{(1-q)r+q(1-\rho)} = 0.                             
    \end{eqnarray}
Therefore, 
   \begin{eqnarray}\nonumber
                     b_P+\lambda &>& w_A b_T-c_f \ \\ \nonumber
            \text{and}\ \          b_P+\lambda &\leq& w_Ab_T-c_m=w_Ab_T,
   \end{eqnarray}
therefore $w_Ab_T-b_P-c_f \leq \lambda \leq w_Ab_T-b_P$. (iii) Suppose $0<\rho<1$. Then note that  women with male gender types are indifferent between competing and not. Hence, 
   \begin{eqnarray}\nonumber
            b_P+\lambda Pr(\tau_i=f\mid d_i=0)&=&w_Ab_T+\lambda \Pr(\tau_i=f\mid d_i=1) \ \\ \nonumber
                                                &>& w_Ab_T-c_f+\lambda \Pr(\tau_i=f \mid d_i=1),
   \end{eqnarray}
      and thus $r=0$ should be true. Then, 
      \begin{eqnarray}\nonumber
            b_P+\lambda \frac{1-q}{1-q\rho} = w_Ab_T \Leftrightarrow \rho=\frac{1}{q}(1-\frac{\lambda}{w_A b_T-b_P}(1-q)),
      \end{eqnarray}
     and $\rho$ is in $(0, 1)$ when  $w_Ab_T-b_P<\lambda \leq \frac{w_Ab_T-b_P}{1-q}$. (iv) Suppose $r=0$ and $\rho=0$. Then we have
     \begin{eqnarray}\nonumber
                   \Pr(\tau_i=f\mid d_i=0)= 1-q.
     \end{eqnarray}
     Off-equilibrium path $\Pr(\tau_i=f\mid d_i=1)$ is assumed to be zero since $\tau_i=f$ is eliminated by the D1 criterion. Therefore, 
     \begin{eqnarray}\nonumber
                 b_P+\lambda(1-q) \geq w_Ab_T(=w_Ab_T-c_m) \Leftrightarrow \lambda \geq\frac{w_Ab_T-b_P}{1-q}.
                      \end{eqnarray}
      This completes the proof.  QED
      
 \ \\
  $Proof\ of\ Proposition\ 3.$ First of all, we show that $r_T=1$. If $r_T<1$, $a_i=0$ is weakly preferred to $a_i=1$ by $\tau_i=f$ and thus, 
     \begin{eqnarray}\nonumber
             w(b_T+\theta_m)-c_m+\lambda \Pr(\tau_i=f\mid a_i=1, d_i=1) &<& w(b_T+\theta_f)-c_f+\lambda \Pr(\tau_i=f \mid a_i=1, d_i=1) \ \\ \nonumber
             &\leq& wb_T-c_f+\lambda \Pr(\tau_i=f\mid a_i=0, d_i=1) \ \\ \nonumber
             &<& wb_T-c_m+\lambda \Pr(\tau_i=f\mid a_i=0, d_i=1),
     \end{eqnarray} 
 hence, $\rho_T=0$. It follows then, $\Pr(\tau_i=f\mid a_i=0, d_i=1)<1=\Pr(\tau_i=f\mid a_i=1, d_i=1)$ and so, 
 \begin{eqnarray}\nonumber
          w(b_T+\theta_f)-c_f+\lambda \Pr(\tau_i=f\mid a_i=1, d_i=1) > wb_T-c_f+\lambda \Pr(\tau_i=f\mid a_i=0, d_i=1),
 \end{eqnarray}
which is a contradiction. Therefore, $r_T=1$. Secondly, we show that $r=1$. Suppose $r<1$. Given that $r_T=1$, note that $\Pr(\tau_i=f\mid a_i=0, d_i=1)=0<\Pr(\tau_i=f\mid a_i=1, d_i=1)$. \footnote{To be more precise, we apply D1 criterion whenever paths are on off-equilibrium.} Then, 
\begin{eqnarray}\nonumber
               \max_{a_i \in \{0, 1\}}w(b_T+\theta_ma_i)-c_m+\lambda \Pr(\tau_i=f\mid a_i, d_i=1) 
               &<&  w(b_T+\theta_f)-c_f+\lambda \Pr(\tau_i=f\mid  a_i=1, d_i=1) \ \\ \nonumber
               &=& \max_{a_i \in \{0, 1\}}w(b_T+\theta_fa_i)-c_f+\lambda \Pr(\tau_i=f\mid a_i, d_i=1) \ \\ \nonumber
               &\leq& b_P+\lambda \Pr(\tau_i=f\mid d_i=0).
 \end{eqnarray}
Hence we have $\rho=0$. However, it follows then $\Pr(\tau_i=f\mid d_i=0)<1$, $\Pr(\tau_i=f\mid a_i=1, d_i=1)=1$ and thus, 
\begin{eqnarray}\nonumber
                 b_P+\lambda \Pr(\tau_i=f\mid d_i=0)&<&b_P +\lambda\ \\ \nonumber
                                                                   &<& w(b_T+\theta_f)-c_f+\lambda \ \\ \nonumber
                                                                   &=&  \max_{a_i \in \{0, 1\}}w(b_T+\theta_fa_i)-c_f+\lambda \Pr(\tau_i=f\mid a_i, d_i=1),
\end{eqnarray}
which is a contradiction. Therefore, $r=1$. Thirdly, we show that $\rho=1$. Given that $r=1$ and $r_T=1$, this is obvious. Finally, let us solve for $\rho_T$. Suppose $0<\rho_T<1$. Then, 
  \begin{eqnarray}\nonumber
                 wb_T = w(b_T+\theta_m)+\lambda \frac{1-q}{1-q(1-\rho_T)},
  \end{eqnarray}
 must hold. Solving this, we obtain, 
 \begin{eqnarray}\nonumber
                  \rho_T=1-\frac{1}{q}(1+\lambda \frac{1-q}{w\theta_m}),
 \end{eqnarray}
and together with the condition $0<\rho_T<1$, we have, 
    \begin{eqnarray}\nonumber
                   -w\theta_m<\lambda<-\frac{w\theta_m}{1-q}.
    \end{eqnarray}
This completes the proof. QED

\end{document}